\documentclass[usenatbib]{mn2e}

\usepackage{graphicx}  
\usepackage{float}
\usepackage{amssymb}   
\usepackage{amsmath}
\usepackage{color}

\newcommand{\hmpc}{\,$h^{-1}$\,Mpc}
\newcommand{\zobov}{{\scshape zobov}}
\newcommand{\origami}{{\scshape origami}}
\newcommand{\camb}{{\scshape camb}}
\newcommand{\avg}[1]{\left\langle{#1}\right\rangle}

\newcommand{\reff}{r_{\rm eff}}
\newcommand{\textwarning}{}

\definecolor{ForestGreen}{rgb}{0.3,0.7,0.3}

\title[Shallower Voids with Warm Dark Matter]{Warmth Elevating the Depths:\\ Shallower Voids with Warm Dark Matter}
\author[Lin F.\ Yang et al.]{Lin F.\ Yang$^{1}$\thanks{E-mail: lyang@pha.jhu.edu}, 
Mark C.\ Neyrinck$^{1}$\thanks{E-mail: neyrinck@pha.jhu.edu},
Miguel A.\ Arag\'on-Calvo$^{2}$,
Bridget Falck$^{3}$,
\and Joseph Silk $^{1,4,5}$\\
$^{1}$Department of  Physics \& Astronomy, The Johns Hopkins University, 3400 N Charles Street, Baltimore, MD 21218, USA\\
$^{2}$Department of Physics and Astronomy, University of California, Riverside, CA 92521, USA\\
$^{3}$Institute of Cosmology and Gravitation, University of Portsmouth, Dennis Sciama Building, Burnaby Rd, Portsmouth PO1 3FX, UK\\
$^{4}$Institut d'Astrophysique de Paris- 98 bis boulevard Arago-75014 Paris, France\\
$^{5}$Beecroft Institute of Particle Astrophysics and Cosmology, Department of Physics, University of Oxford, \\ Denys Wilkinson Building, 1 Keble Road, Oxford OX1 3RH, UK
}

\begin{document}
\date{}

\pagerange{\pageref{page0}--\pageref{pageend}} \pubyear{2015}

\maketitle

\begin{abstract}
Warm dark matter (WDM) has been proposed as an alternative to cold dark matter (CDM), to resolve issues such as the apparent lack of satellites around the Milky Way. Even if WDM is not the answer to observational issues, it is essential to constrain the nature of the dark matter. The effect of WDM on haloes has been extensively studied, but the small-scale initial smoothing in WDM also affects the present-day cosmic web and voids. It suppresses the cosmic ``sub-web" inside voids, and the formation of both void haloes and subvoids. 
In $N$-body simulations run with different assumed WDM masses, we identify voids with the \zobov\ algorithm, and cosmic-web components with the \origami\ algorithm. As dark-matter warmth increases {\textwarning (i.e., particle mass decreases), void density minima grow shallower, while void edges change little. Also, the number of subvoids decreases. The density field in voids is particularly insensitive to baryonic physics, so if void density profiles and minima could be measured observationally, they would offer a valuable probe of the nature of dark matter.} Furthermore, filaments and walls become cleaner, as the substructures in between have been smoothed out; this leads to a clear, mid-range peak in the density PDF.
\end{abstract}

\label{page0}


\section{Introduction}

There is overwhelming evidence \citep[e.g.][]{frenk2012dark} for the existence of dark matter (DM), whose nature is still unknown.  Although many direct or indirect detection experiments (e.g. \cite{PhysRevLett.112.091303, PhysRevLett.111.251301, aprile2013xenon1t, collaboration2013dark}) have occurred and are ongoing, no conclusive result has been reported.  

It has been long known \citep{peebles1980large, bertone2005particle} that the matter of the universe is dominated by DM. For structure formation, the velocity distribution of DM particles plays an important role. In the standard $\Lambda$CDM model, the DM is assumed to be entirely cold from the standpoint of structure formation. The velocity dispersion is negligible at the era of matter-radiation equality ($t_{\rm eq}$), and structure formation proceeds in a bottom-up fashion. Smaller structures form at first, then larger ones. This model has only a few parameters, that have been determined with high precision.  However, several problems remain unsolved on sub-galactic scales. First, the missing satellite problem: simple arguments applied to CDM-only simulations imply that thousands or hundreds of dwarf galaxies are expected in the local group and halo of the Milky Way,  however only of order 10 of them were found \citep{moore1999dark, mateo1998}. Second, CDM predicts concentrated density profiles in the central region, e.g. $r^{-1}$ in the NFW \citep{1997ApJ...490..493N} profile, whereas many studies of galaxy rotation curves have concluded that the density approaches a constant in the core \citep{moore1999cold, ghigna2000density}. Third, the number of dwarf galaxies expected in local voids may be less than a CDM model would predict \citep{peebles2001void}.

Although better modelling of hydrodynamics and feedback processes may solve these problems \citep[e.g.\ ][]{hoeft2006dwarf}, changing the DM itself could also help resolve some of the issues.  Warm dark matter (WDM) has been an attractive alternative since the 1980s \citep[e.g.\ ][]{schaeffer1988cold}. 
Recently, the WDM model has received some interest 
since it can reproduce all the successful CDM results on large scales, but also solve some small-scale issues. The key feature separating WDM  from CDM models is the lack of initial small-scale fluctuations.  WDM has slightly larger velocity dispersion at $t_{\rm eq}$, giving a smoothing of initial fluctuations at a free-streaming length determined by the WDM particle mass. From particle physics, the originally favored WDM candidate was a gravitino \citep[e.g.][]{moroi1993cosmological}; more recently, a sterile neutrino \citep{boyarsky2009lyman} has seen attention.  Both theoretical and numerical studies (e.g.  \cite{bode2001halo}) have explored WDM models, and observational constraints have been put on the mass of WDM particles. E.g. \cite{viel2013cold} give a lower limit of $m_{X}=3.3$ keV from Lyman-$\alpha$ forest data (HIRES data). Other independent studies (e.g. \cite{miranda2007constraining}) also give consistent limits. 

While previous studies on WDM mainly focused on the formation of halos or other dense structures, there has been some work investigating the cosmic web itself. \citet{SchneiderEtal2012}  studied voids in a WDM scenario, but focus on the halo population within them, finding that voids are emptier (of haloes and substructure) in WDM.  Below, we study the dark-matter density itself, which follows the opposite trend, growing in density in WDM. \citet{ReedEtal2014} also studied the large-scale-structure traced out by galaxies in a WDM scenario, including a study of galaxy environment. They found that WDM makes very little difference in the usual observables of the galaxy population, when using subhalo abundance matching (SHAM) to identify galaxies. That is, they found that the subhaloes in the dark matter, above a mass threshold where halo formation is not substantially disrupted from the loss of power in WDM, are arranged in nearly the same way in WDM and CDM. Hydrodynamic simulations that include star formation indicate that stars may form in filaments instead of haloes if the dark matter is quite warm \citep{GaoTheuns2007,GaoEtal2014}, an issue related to the low ``complexity''  of dark-matter halo structure in WDM \citep{Neyrinck2014info}.

Voids are large underdense regions, occupying the majority of the volume of the Universe, and are valuable cosmological probes.  For example, via the Alcock-Paczynski test, voids serve as a powerful tool to detect the expansion history of the universe (e.g.\ \cite{ryden1995measuring, lavaux2012precision}). \cite{clampitt2013voids, li2012haloes} have also proposed voids as a probe of   modified gravity.  The abundance of voids may be sensitive to  initial conditions \citep{kamionkowski2009void}, hence voids may serve as probes of the early universe. Previous work has also looked at voids in the context of WDM. \citet{tikhonov2009sizes} measured the abundance of mini-voids, which become scarcer in a WDM model.
Recently, \citet{clampitt2014lensing} detected void lensing at a significance of 13$\sigma$, raising hopes for void density-profile measurements using lensing. The few-parameter ``universal'' form for void density profiles that \citet{hamaus2014focus} found will likely help in extracting cosmological information from voids.

In this paper, we study how properties of voids in the cosmic web change in a WDM scenario, with different initial power-spectrum attenuations corresponding to different WDM masses. We analyze these simulations with the \zobov\ void-finder \citep{neyrinck2008zobov} and the \origami\ \citep{FalckEtal2012}, filament, wall and halo classifier.  The paper is laid out as follows. In section 2, we introduce our warm dark matter N-body  simulations. In section 3, we analyze the full cosmic web of DM in a WDM scenario. In section 4, we introduce our  void detection methods and shows the void statistical properties.  In section 5, we show the distinct features of void density profiles for different DM settings. We give our conclusions and discussion in section 6.

\section{ Simulations}
We simulate  both CDM and WDM  using  the {\scshape gadget}-2 \citep{2005MNRAS.364.1105S} code. 
We use the Zel'dovich approximation \citep{zeldovich1970} to impart initial displacements and velocities at  redshift $z=127$ to particles on the initial lattice of $512^3$ particles in a periodic box of size 100\hmpc.
{\textwarning The initial power spectrum was generated with} the \camb~code \citep{lewis2000camb}, using vanilla $\Lambda$(C)DM cosmological parameters ($h=0.7$, $\Omega_{\rm M}=0.3$, $\Omega_{\rm \Lambda}=0.7$, $\Omega_{\rm b}=0.045$, $\sigma_8=0.83$, $n_s=0.96$). 

To incorporate the effect of a thermally produced relic WDM particle, we apply the following fitting formula to the transfer functions \citep{bode2001halo},
\begin{equation}
\label{eqn:fitt}
T_\mathrm{WDM}=T_\mathrm{CDM}(k)[1+(\alpha k)^2]^{-5.0},
\end{equation}
where the cutoff scale, 
\begin{align}
\alpha =0.05\left(\frac{\Omega_m}{0.4}\right)^{0.15}\left(\frac{h}{0.65}\right)^{1.3}
	\left(\frac{m_\mathrm{dm}}{1\mathrm{keV}}\right)^{-1.15}&\left(\frac{1.5}{g_X}\right)^{0.29}.\label{eqn:bodetransfer}\\
	&h^{-1}\mathrm{Mpc},\nonumber
\end{align}
Here, $m_\mathrm{dm}$ is the mass of the WDM particle (or the effective sterile neutrino);  $g_X$ is the number of degrees of freedom that the WDM particle contributes to the number density (in our case, $3/2$). In our set of simulations, we applied $\alpha=0$, 0.05, 0.1 and $0.2$ \hmpc, corresponding first to CDM, and then to WDM particle masses $1.4$, $0.8$ and $0.4$ keV. 
{\textwarning Note that these masses of sterile neutrinos are disfavoured by Lyman-$\alpha$ forest data \citep{viel2013cold}, but we adopt them to show the effect of WDM without great computational cost. }

Only in the most extreme case of $0.2$ \hmpc\ is $\alpha$ comparable to the interparticle separation. The reason that differences show up even when $\alpha$ is below this scale is the broad shape of the attenuation described in Eq.\ (\ref{eqn:bodetransfer}).   \citet{bode2001halo} define a perhaps more meaningful ``half-mode'' scale radius $R_s$, via $T(\pi/R_s)=1/2$; this quantity is $\sim 6\alpha$. We hold $\sigma_8$ fixed in the linear power spectrum.  This changes the large-scale amplitude, but very slightly since the smoothing kernel acts on scales well below 8\hmpc.

\begin{figure}
    \centering
     \includegraphics[width=\columnwidth]{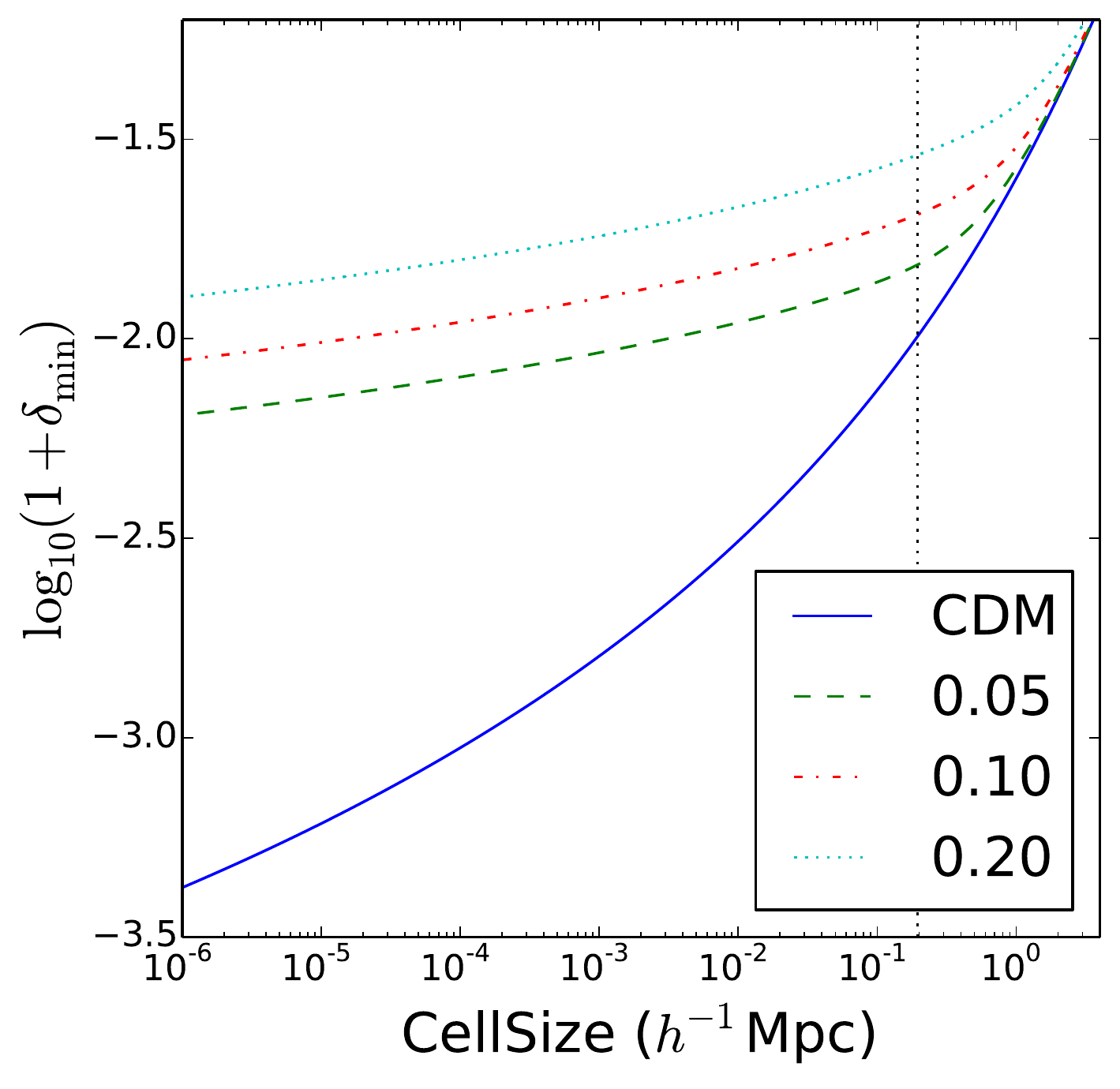}
    \caption{The minimum density expected in a (100\hmpc)$^3$ volume, as a function of simulation resolution. The four curves show the four cases investigated below: the WDM cut-off scale $\alpha=0$ (CDM), 0.05, 0.1, and 0.2\hmpc. The cell size is the initial comoving interparticle separation. A dashed line shows the cell size for the resolution used in the simulations used in this work.}
    \label{fig:deltares}
 \end{figure}

{\textwarning  We emphasize that the differences between CDM and the WDM models would increase with the mass resolution, because even with ``$\alpha=0$'',  there is a cutoff in the initial power spectrum from the interparticle spacing of 0.2 \hmpc. To illustrate this, we show in Fig.\ \ref{fig:deltares} the expected minimum density, as a function of resolution, in a 100 \hmpc\ box, for CDM and WDM models. We estimate this as follows. As shown below in Fig.\ \ref{fig:origawdmi}, the spherical-collapse limit \citep{Bernardeau1994,ProtogerosScherrer1997,Neyrinck2013}
\begin{equation}
\delta_{\rm sc}+1 = \left[1-(2/3) \delta_{\rm lin}\right]^{-3/2}
    \label{fig:deltasc}
\end{equation}
accurately gives the transformation from linear to non-linear density on the low-density tail. The lowest linear-theory density in the box will be the expected minimum value of a Gaussian in a sample of $N_{\mathrm{cells}}=[(100$\hmpc$)/c]^3$ cells, 
where $c$ is the cell size. This is
\begin{equation}
\delta_{\rm lin}^{\rm min}=-\sigma(c)\sqrt{2}{\rm erf}^{-1}(2/N_{\rm cells})
\end{equation}
where $\sigma(c)$ is the linear-theory density dispersion in cells of size $c$. (We approximate $\sigma(c)$ with a spherical top-hat kernel of the same volume as a cubic cell.) As $c$ decreases, the WDM and CDM curves diverge, because $\sigma(c)$ increases in CDM, but not in WDM. The WDM curves do continue to decrease slightly, however, because the number of cells increases.
}

We have not included the thermal velocity kicks to the individual particles to our simulation, for two reasons.  First, particles in the simulation are averages over a statistical ensemble of particles, so it is unclear how to implement the thermal velocity in the initial conditions. Also, this thermal velocity would be negligible for our results. For the  $\alpha$ values we use, the RMS velocity distribution is of order $1$ km/s \citep{angulo2013warm}, while typical velocities of particles inside a void are of order $50$ km/s at radius around 0.5$r_v$ \citep{Hamaus:2014vy}. We do not employ a method to prevent spurious fragmentation in filaments \citep{WangWhite2007}, because we focus on voids instead of haloes. {\textwarning Spurious fragmentation should only upscatter particles in morphological type, i.e., turning wall particles into filament or halo particles, and filament particles into halo particles. This artificial fragmentation happens in the presence of anisotropy, but should not happen in voids, where even initially anisotropic volume elements grow nearly isotropic with time \citep{Icke1984}.}

\section{The Cosmic Web in a WDM Scenario}

  \begin{figure*}
    \begin{center}
     \includegraphics[width=1.0\textwidth]{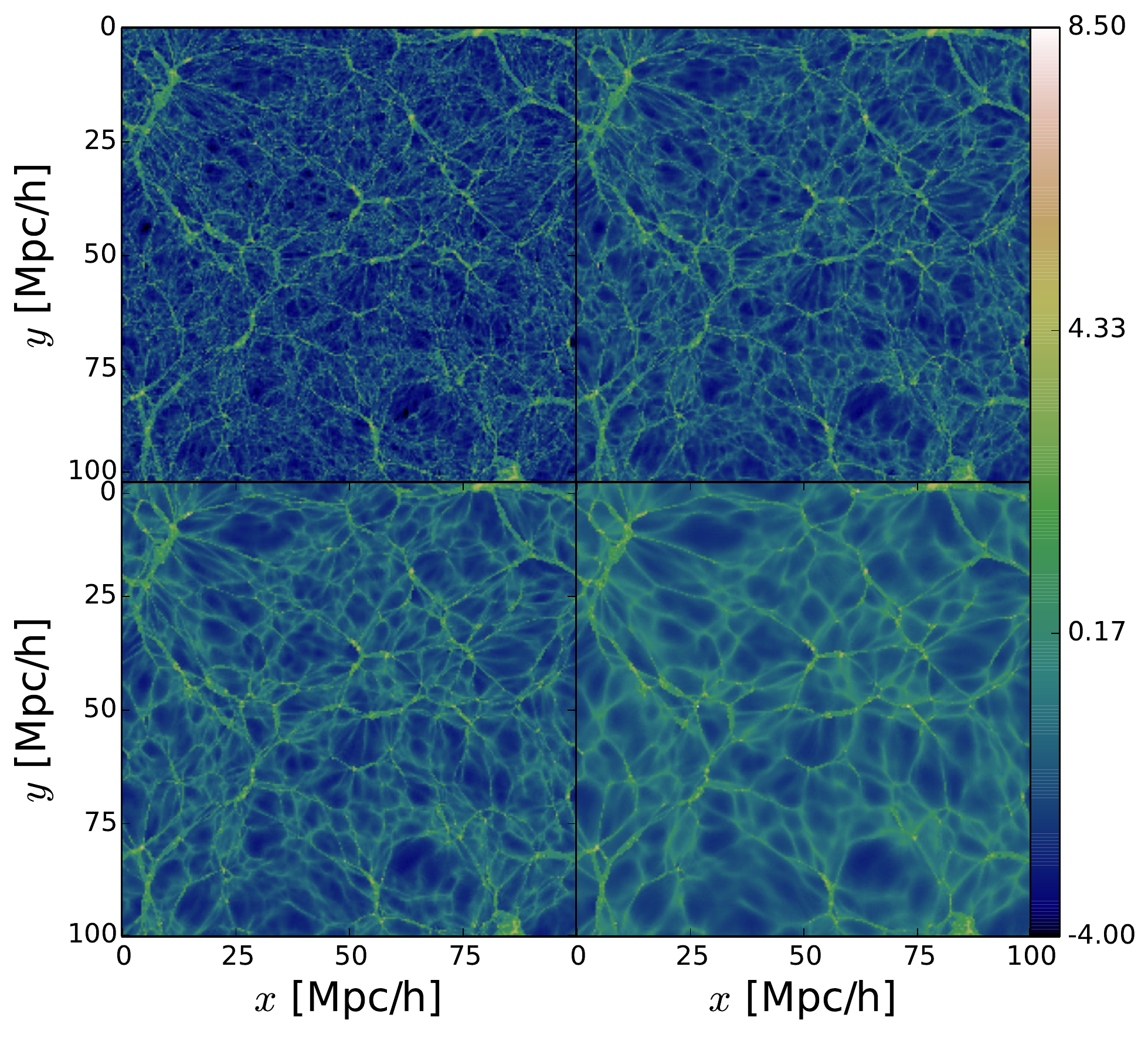}
     \end{center}  
    \caption{LTFE density field slices, showing $\ln(1+\delta)$.  From top left to bottom right: CDM, WDM with $\alpha=0.05$ \hmpc, WDM with $\alpha=0.1$ \hmpc and WDM with $\alpha=0.2$\hmpc. }
    \label{fig:densfield}
 \end{figure*}
Fig.\ \ref{fig:densfield} shows a slice of an LTFE \citep[Lagrangian Tessellation Field Estimator;][]{abel2012ltfe} density field of the simulation. This estimator makes use of the the fact that, under only gravity, the 3D manifold of DM particles evolve in phase space without tearing, conserve phase space  volume and preserve connectivity of nearby points.  Hence the sheets (or streams) formed by the initial grids are assumed to remain at constant mass in the final snapshot. Using the initial grid position (or the Lagrangian coordinates) of each particle, the density of each stream could be calculated. We implemented an OpenGL code of LTFE to estimate the density field in a time efficient way. The differences of density fields are clear: small-scale structures are smoothed out in the WDM simulations.

\begin{figure*}
  \begin{minipage}{\textwidth}
    \begin{center}
    \includegraphics[width=\textwidth]{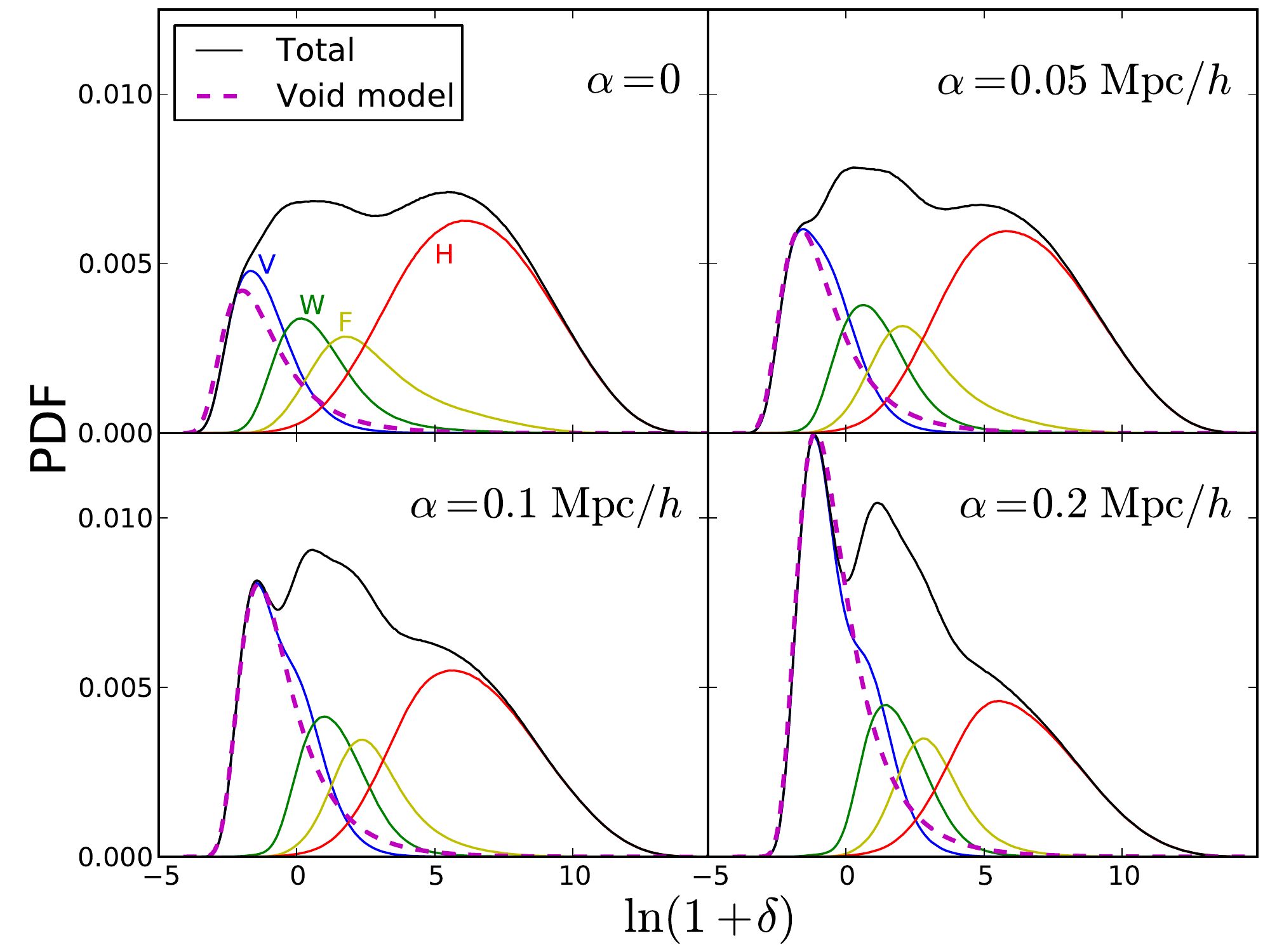}
  \end{center}  
  \caption{PDFs of particle densities for the four simulations.  A distinctive peak arises from filament and wall particles at middling densities in a WDM scenario. The V, W, F, and H curves add up to the total: they separate out void, filament, wall, and halo particles, with crossings along 0, 1, 2, and 3 orthogonal axes. $\alpha=0$ corresponds to CDM; as $\alpha$ increases, the WDM becomes warmer.  In this mass-weighted PDF, each particle (Lagrangian element of initial spacing 0.2\hmpc) enters once. The dashed magenta curve shows the expression in Eq.\ (\ref{eqn:pdfA}).}
  \label{fig:origawdmi}
  \end{minipage}
\end{figure*}

Fig.\ \ref{fig:origawdmi} shows mass-weighted 1-point PDFs (probability density functions) at $z=0$ from the simulations, for each $\alpha$. We measured the density at each particle using the Voronoi Tessellation Field Estimator \citep[VTFE]{svdw,vdws}. In the VTFE, each particle occupies a Voronoi cell, a locus of space closer to that particle than to any other particle. The density $\delta_{\rm VTFE}+1=\avg{V}/V$ at a particle is set by the volume $V$ of its cell.  This density measure is in a sense Lagrangian, but only strictly so without multi-streaming.

At $\alpha=0$, this mass-weighted PDF shows two clear peaks, noted by \citet{neyrinck2008zobov}. It was already clear that the higher-density peak, a roughly lognormal peak at $\delta\approx e^{6-7}\approx1000$, comes from halo particles. \citet{FalckEtal2012} firmly established this halo origin by classifying particles with the \origami\ algorithm, into void (single-stream), wall, filament, and halo morphologies. This algorithm counts the number of orthogonal axes along which a particle has been crossed by any other particle, comparing the initial and final conditions.

For $\alpha=0$, the lumpy shape of the total PDF at low densities already suggests that there may be more than 2 components. As $\alpha$ increases, however, an intermediate wall+filament peak becomes unmistakable: the visual impression from the density-field maps that a greater fraction of the matter is in walls and filaments is obvious in the total PDF as well. Again, the \origami\ classification confirms this picture. As the WDM mass increases, particles move from low to high densities and morphologies, through the different peaks.  A similar effect happens as a function of simulation resolution in CDM: at higher resolution (smaller interparticle separation), the fraction of halo particles increases and fraction of void particles decreases. However, no obvious intermediate wall+filament peak appears in the total CDM PDF \citep{FalckNeyrinck2015}. The fractions of particles in walls and filaments remain about constant, with most of the change in particle morphologies appearing in the void and halo peaks
. There is a change in the mean log-densities of the wall and filament peaks, however; while a substantial fraction of wall particles have $\delta<0$ with CDM, note that only the end of the wall tail has $\delta<0$ when $\alpha=0.2$.

We note that this complicated PDF shape is likely poorly constrained by its first few moments, certainly not the moments of the overdensity $\delta$, and likely not even the log-density $\ln(\delta+1)$, the $x$-axis of the plot \citep[e.g.][]{Carron2011,CarronNeyrinck2012}. This is a case in which analyzing the  instead of the first few moments would be prudent \citep[e.g.][]{neyrinck2014transformationally,HillEtal2015}.

Also plotted is a successful analytic expression for the distribution of void-particle densities \citep{ProtogerosScherrer1997,Neyrinck2013}, derived from a low-$\Omega_{\rm M}$ limit (especially valid for voids) to the evolution of an average mass element \citep{Bernardeau1994}. The PDF of the log-density $A\equiv\ln(1+\delta)$ is
\begin{equation}
  P(A)=f_{\rm void}\frac{\exp\left[-\frac{2}{3}A-\frac{(3/2)^2}{2 \sigma^2} \left(e^{-(2/3)A}-1\right)^2\right]}
{\sqrt{2\pi\sigma^2}},
  \label{eqn:pdfA}
\end{equation}
where $\sigma^2$ is the linearly-extrapolated initial variance in cells of size the initial interparticle spacing. $f_{\rm void}$ is the fraction of void, single-stream particles, as measured by \origami. $\sigma^2$ is the variance in spheres of radius $L/N/(4\pi/3)^{1/3}$ (where $L=100$\hmpc\ is the box size, and $N=512$) as calculated from a \camb\ linear power spectrum at $z=0$, truncating the power spectrum to zero at $k>\pi/(L/N)$.

 \begin{figure}
     \begin{center}
    \includegraphics[width=0.9\columnwidth]{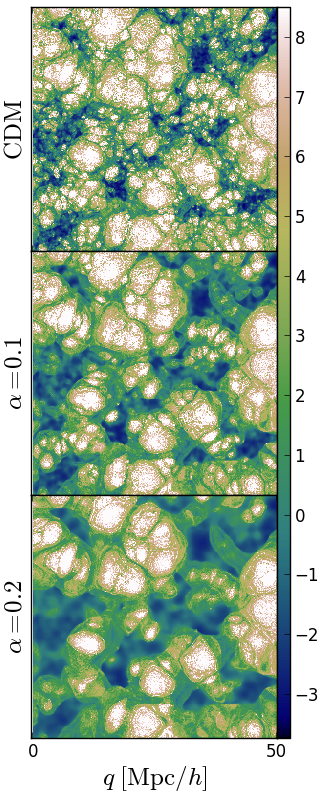}
  \end{center}  
    \caption{$z=0$ Voronoi particle densities $\ln(1+\delta)$ on a 2D Lagrangian sheet, with one particle per $0.2$\hmpc\ comoving Lagrangian pixel. Each panel shows a 2D, $256^2$ slice, a quadrant of a full $512^2$ slice.}
    \label{fig:wdm_vol}
 \end{figure}

 \begin{figure}
     \begin{center}
    \includegraphics[width=0.9\columnwidth]{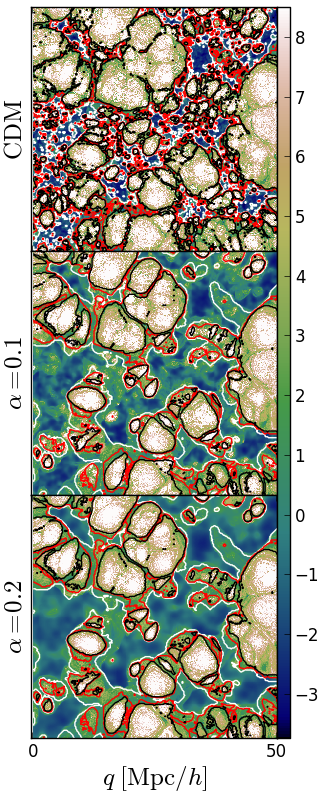}
  \end{center}  
    \caption{Same as Fig.\ \ref{fig:wdm_vol}, with \origami\ morphologies added: black, red, and white contours separate void, wall, filament, and halo particle morphologies.}
    \label{fig:wdm_vol_origami}
 \end{figure}

In Figs.\ \ref{fig:wdm_vol} and \ref{fig:wdm_vol_origami}, Lagrangian-space density maps show where these various density regimes appear in the cosmic web. Here, each pixel represents a particle, arranged on its initial lattice. In Fig.\ \ref{fig:wdm_vol_origami}, wart-like blobs within black contours are haloes; these contract substantially in the mapping to comoving $z=0$ Eulerian space. The regions within white contours are void regions, which expand in comoving coordinates and come to fill most of the space; see \cite{FalckEtal2012} for more detail and an alternative plotting method.

Regarding the topology of the void region, an increase in $\alpha$ decreases the amount of stream-crossing. At high $\alpha$, even though filaments and walls become more evident visually in the density field, the decreased stream-crossing makes the percolation of the void region even more obvious than in CDM \citep{FalckNeyrinck2015}. Even in the 2D Lagrangian slice in Fig.\ \ref{fig:wdm_vol_origami}, the void region obviously does pinch off into idealized convex voids \citep[e.g.][]{IckeVdw1991,Neyrinck2014origami}.

\section{Void Detection and Properties}

\zobov\ first uses a Voronoi tessellation to get the density of each particle. After that, it uses each local density minimum as a seed and groups other particles around it using the watershed algorithm, forming a ``zone'', regions with a density  and a ridge. These zones are combined into larger parent voids using essentially another application of the watershed algorithm, giving a hierarchy of subvoids. 

In the default ``parameter-free" algorithm output, the whole field  is a single large super-void with many levels of sub-voids, which is difficult to use directly. We therefore required the zones added to a void to have core density less than $\bar{\rho}$, the mean density of the simulation. \zobov\ measures the statistical significance of voids, compared to a Poisson process.  The probability a void is real depends on the density contrast, defined as the ratio of the minimum density on the ridge separating the void from another void to the void's minimum density. 
To focus on voids with low discreteness effects, we analyze voids with significance larger than $3\sigma$ according to this density contrast criterion and measure their properties. We found about $7600$, $6700$, $3900$, and $1400$ voids for the 4 simulations. Consistently with the results of  \citet{tikhonov2009sizes}, the number of voids decreases as the cut-off scale increases.  Compared to \cite{Hamaus:2014vy}, our much higher mass resolution (compared to their sparse-sampling to better approximate a galaxy sample) allows much smaller voids to be detected in the matter field. We acknowledge that these smallest structures would likely not show up in a galaxy survey unless it was extremely deep. Even then, small voids may be smeared out by redshift-space distortions. However, some of these small (sub)voids would be in low-density regions, with smaller redshift-space distortions smearing them out. And also, we emphasize that many of these small void cores, at low sampling, would also likely be centers of larger, parent voids, that are not necessarily included in our pruned catalog.

 \begin{figure}
    \centering
     \includegraphics[width=\columnwidth]{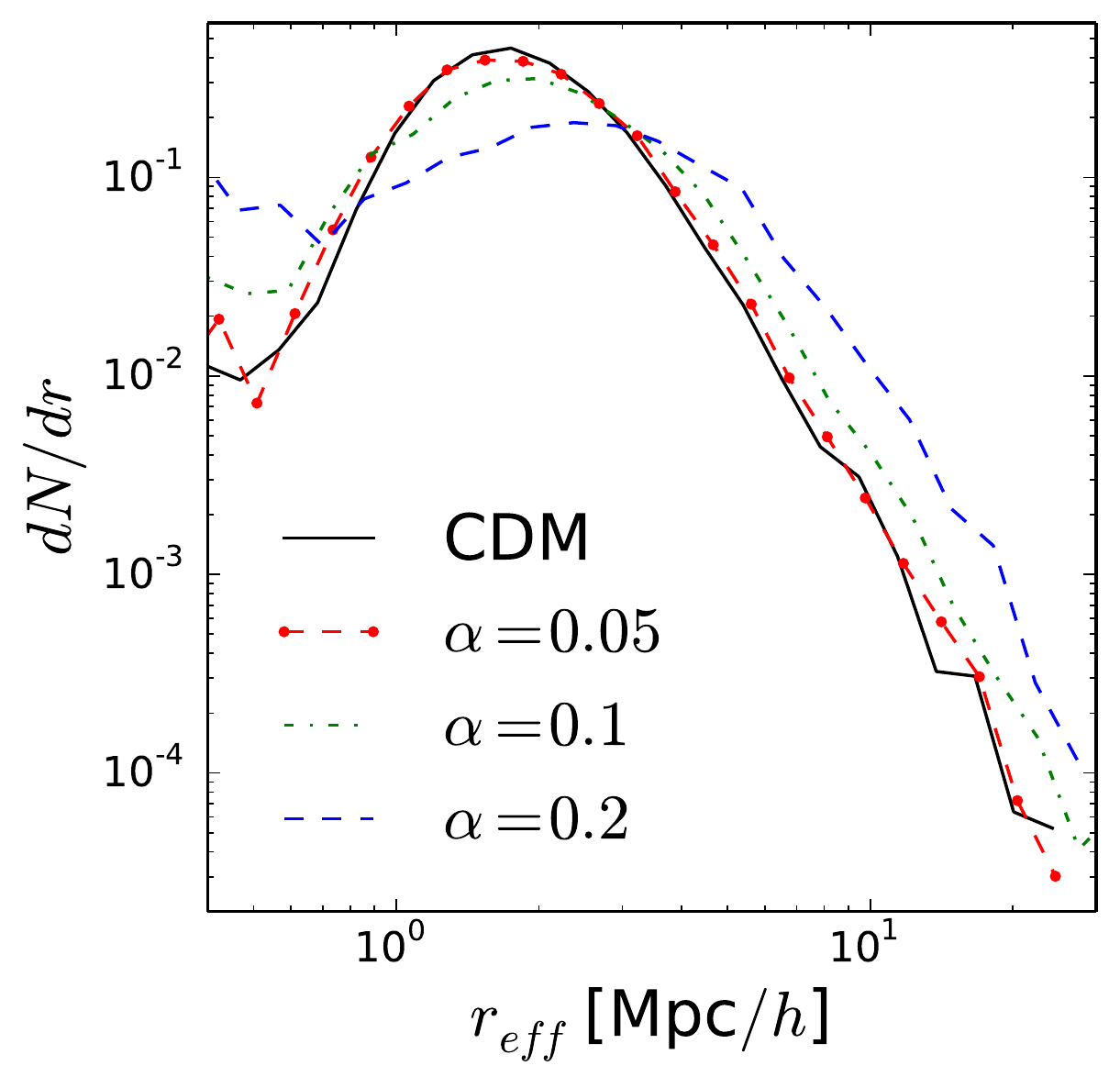}
    \caption{The void radius $\reff$ distribution for different dark matter models. }
    \label{fig:voidreff}
 \end{figure}
 \begin{figure}
    \centering
     \includegraphics[width=\columnwidth]{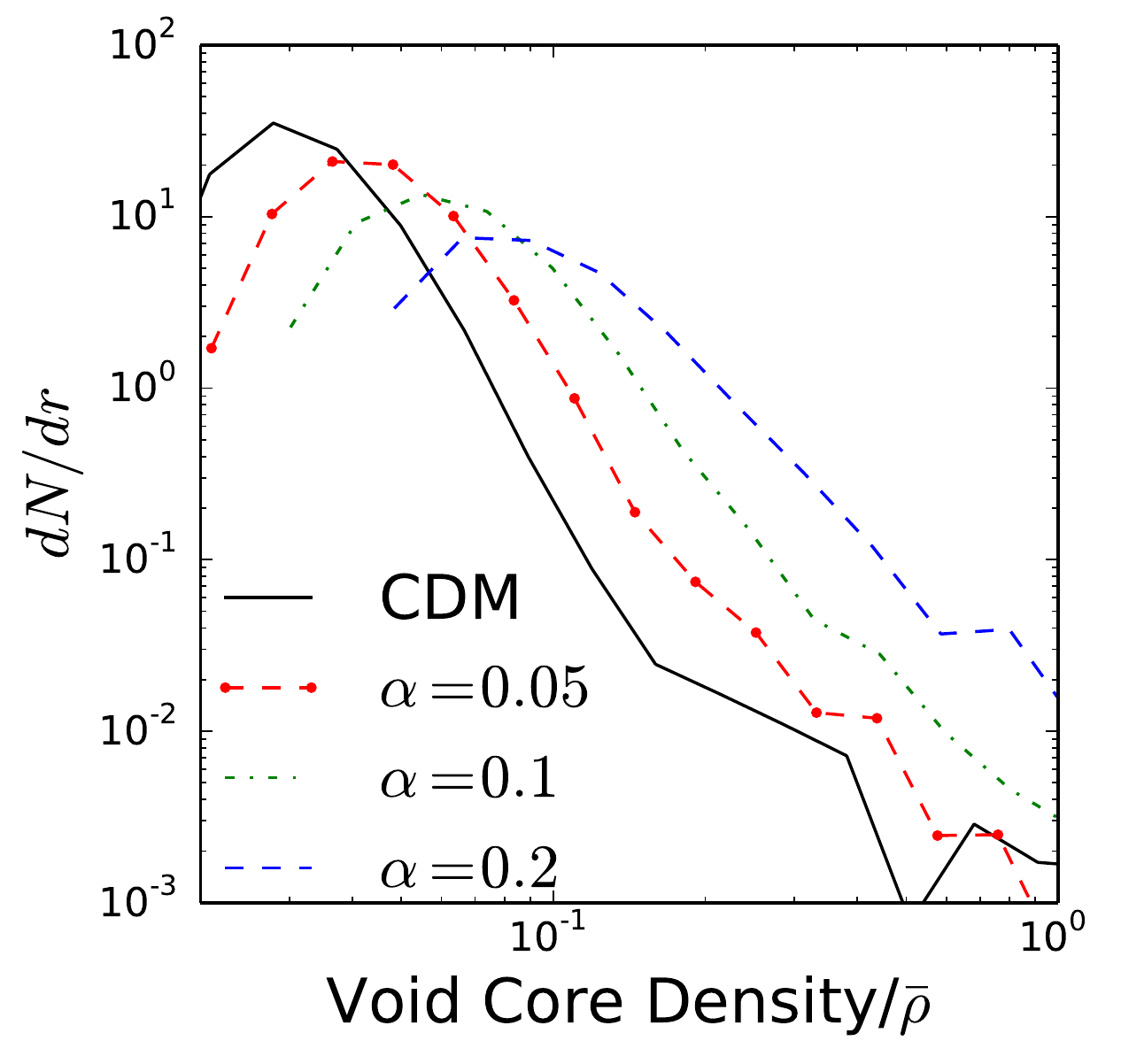}
    \caption{The void core density distribution for different dark matter models.}
    \label{fig:voidcoredens}
 \end{figure}

{Comparing to  \citet{2013MNRAS.428.3409A, 2010MNRAS.404L..89A}, which considered the hierarchical structure of voids in different levels of smoothing, we do not explicitly take out the sub-voids in each level of the hierarchy but use their effective sizes  to characterize them. Apparently, larger voids show up in lower levels of the hierarchy tree.}
Fig.\ \ref{fig:voidreff} shows distributions of void effective radius, $\reff$, defined via $V_r = 4\pi/3\reff^3$. The size of the voids in our simulations peaks roughly at $2$ \hmpc, and shifts slightly outward at high $\alpha$ (moving to WDM).
The abundances of voids around the interparticle spacing, $0.2$\hmpc, are very sensitive to discreteness noise. Thus, we truncate each plot on the left at twice the mean particle separation, at $\sim 0.4$\hmpc.

  In the right tails ($\reff \gtrsim 3$ Mpc) of Fig. \ref{fig:voidreff}, there are more large voids in a WDM scenario, but not dramatically so. This may depend more on the void definition than the distribution of core densities, however, since the reported radius of a void obviously depends crucially on where its boundary is drawn.  {While at radius $\gtrsim 1$Mpc, the abundance depression on the  left  and increasing at the right of the figure clearly shows the effect that sub-structures in larger voids have been suppressed.} Fig.\ \ref{fig:voidcoredens} shows the minimum ``core'' Voronoi density distribution for voids and sub-voids (in units of the mean density $\bar{\rho}$). It shows that the centers of voids become shallower in WDM. This is a simple physical effect: in WDM, the initial density PDF on the scale of the interparticle spacing is narrower than in CDM, because of the small-scale attenuation. This results in a narrower particle density distribution at $z=0$, as shown in Sec.\ 3. In particular, density minima, in the low-density tail, increase in density.

 \begin{figure}
    \centering
     \includegraphics[width=0.45\textwidth]{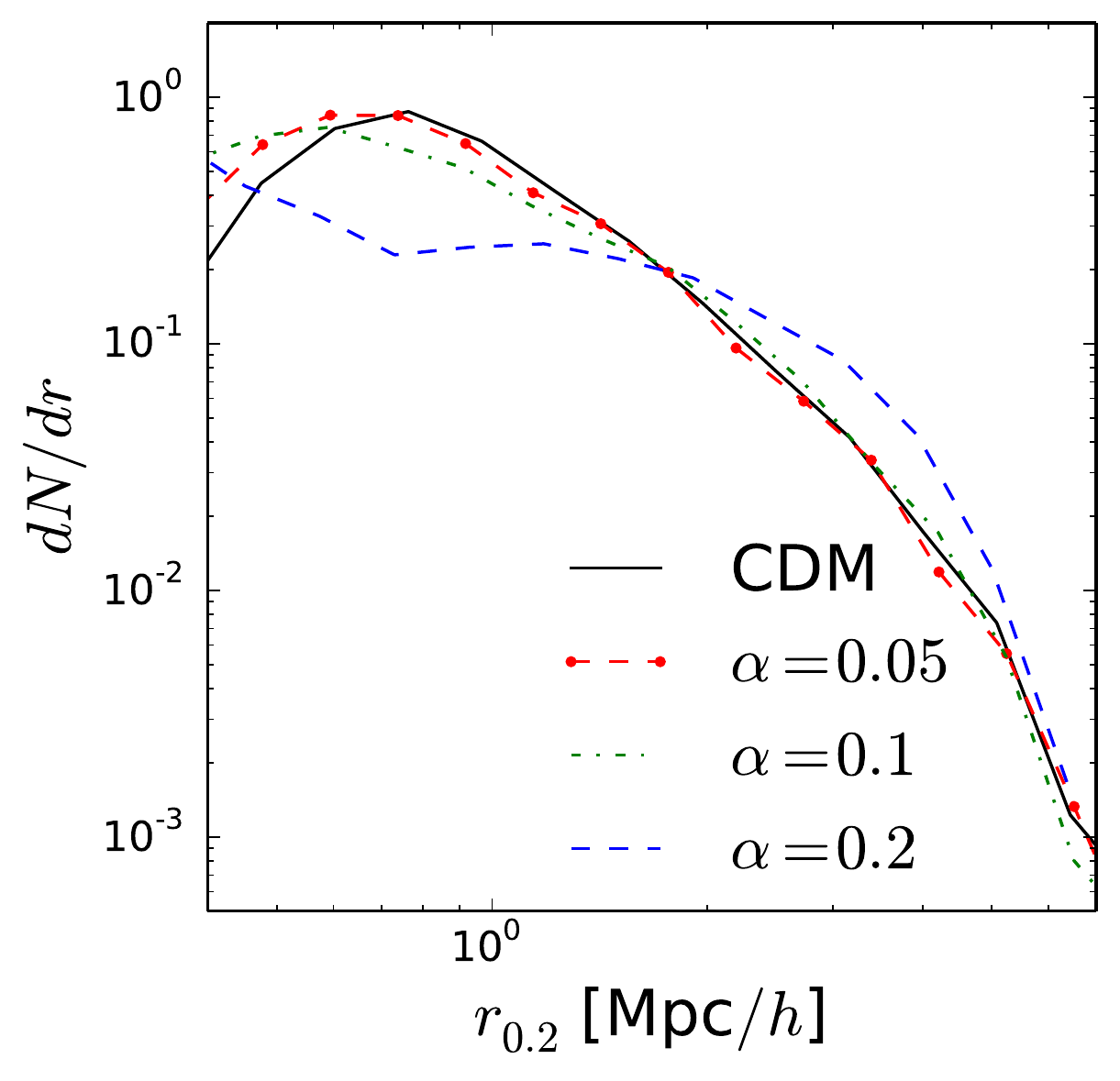}
    \caption{The distribution of  $r_\mathrm{0.2}$ for different dark matter models.}
    \label{fig:r02pdf}
 \end{figure}

\section{Density profiles}

 \begin{figure*}
    \centering
     \includegraphics[width=0.45\textwidth]{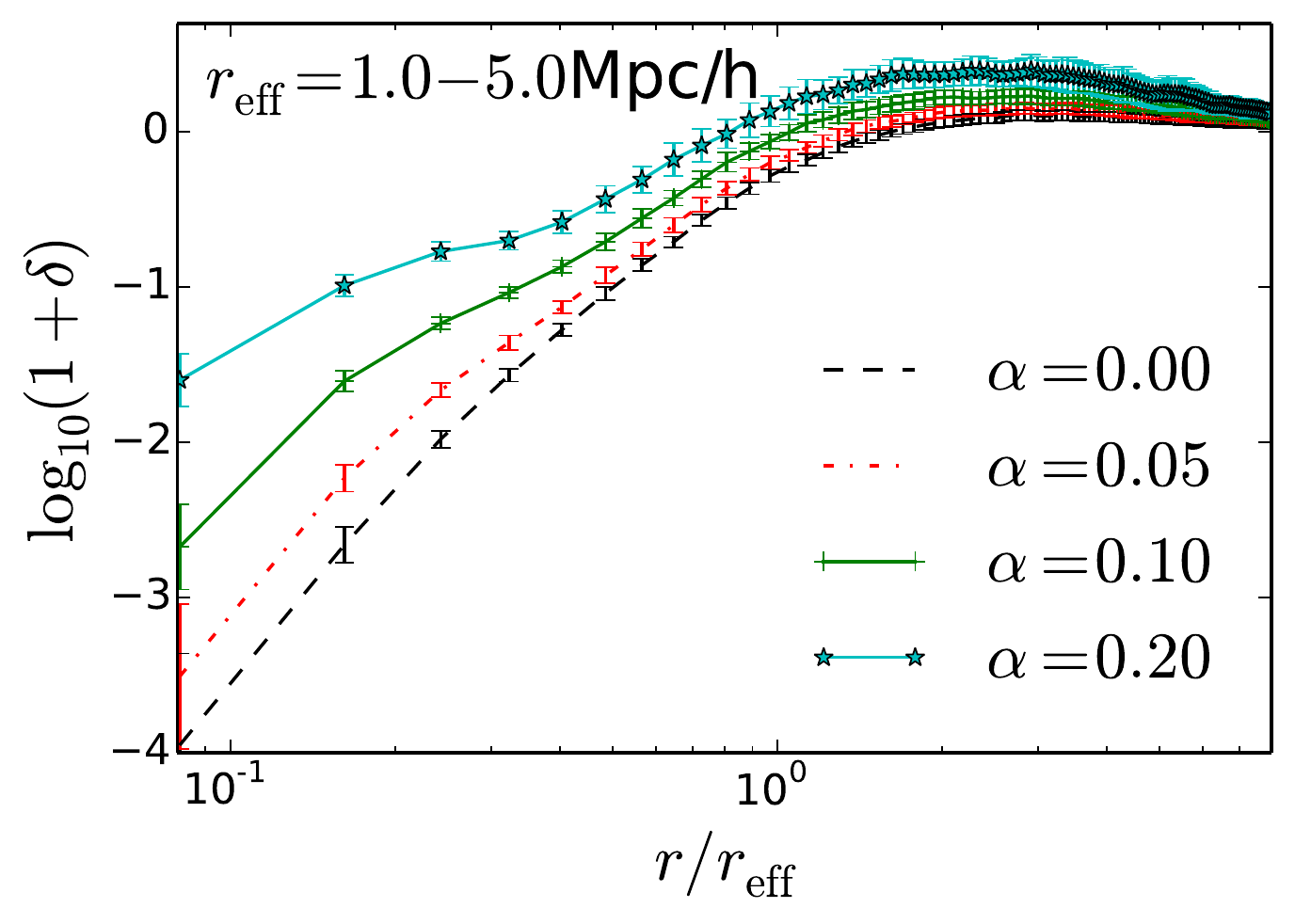}
     \includegraphics[width=0.45\textwidth]{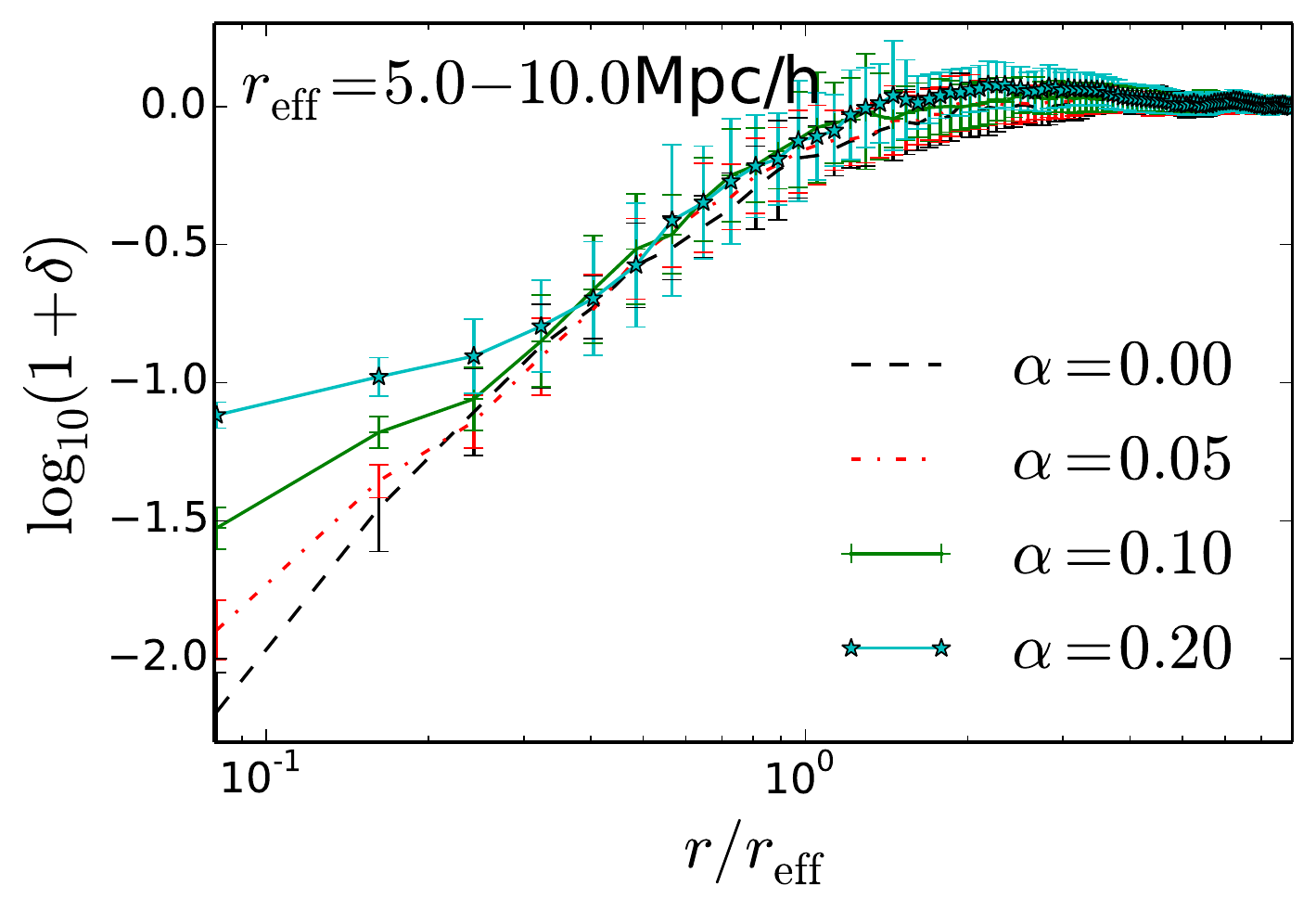}
    \caption{Void density profiles measured and scaled with $\reff$. The origin is the density minimum of each void. The left panel shows voids in the 1-5 \hmpc\ radius bin; the right panel shows results from 5-10 \hmpc\ voids.  Error bars show the $2\sigma$ error, dividing by $\sqrt{N}$, where $N$ is the number of stacked voids.}
    \label{fig:densprof1}
 \end{figure*}

 \begin{figure*}
    \centering
     \includegraphics[width=0.45\textwidth]{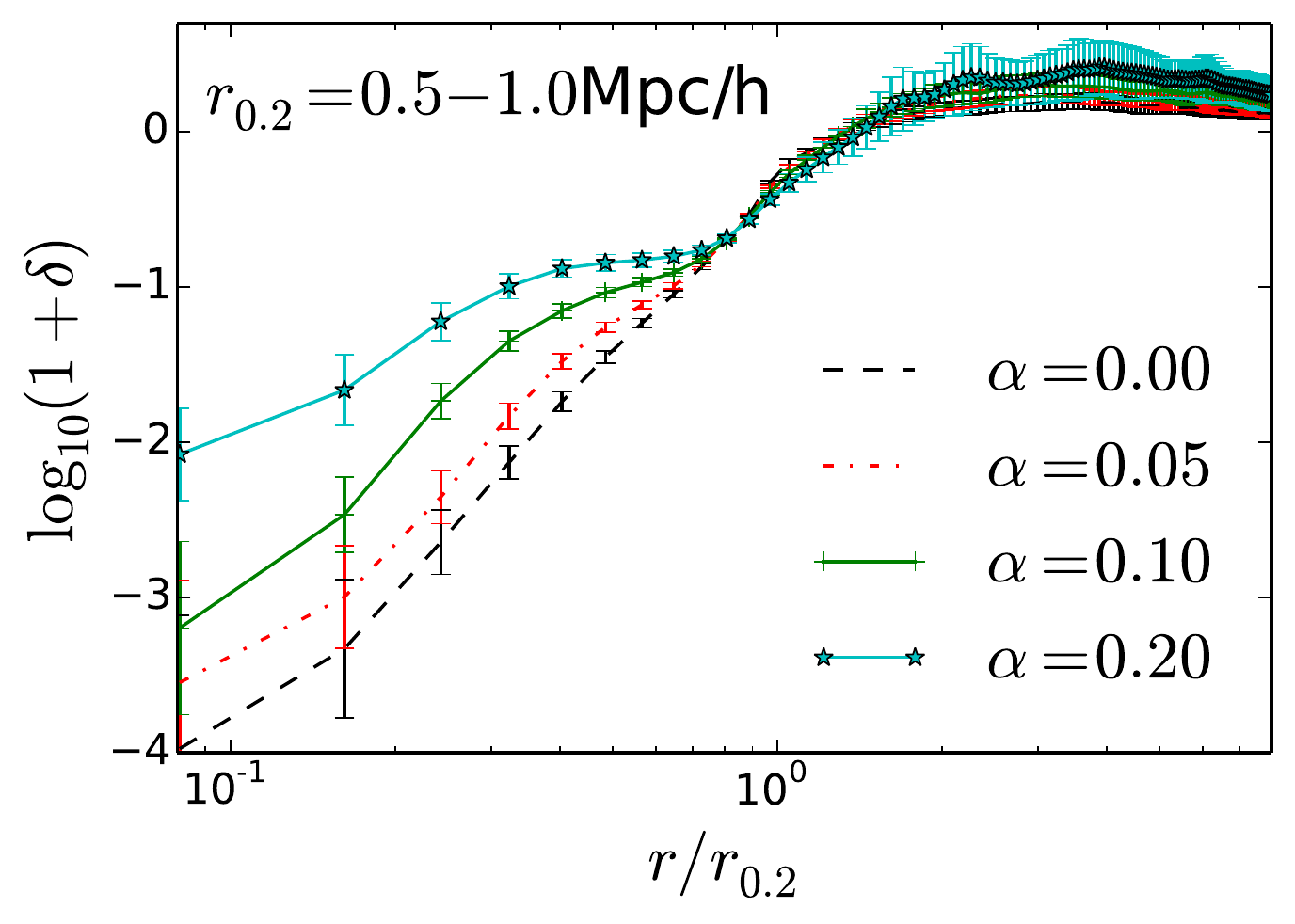}
     \includegraphics[width=0.45\textwidth]{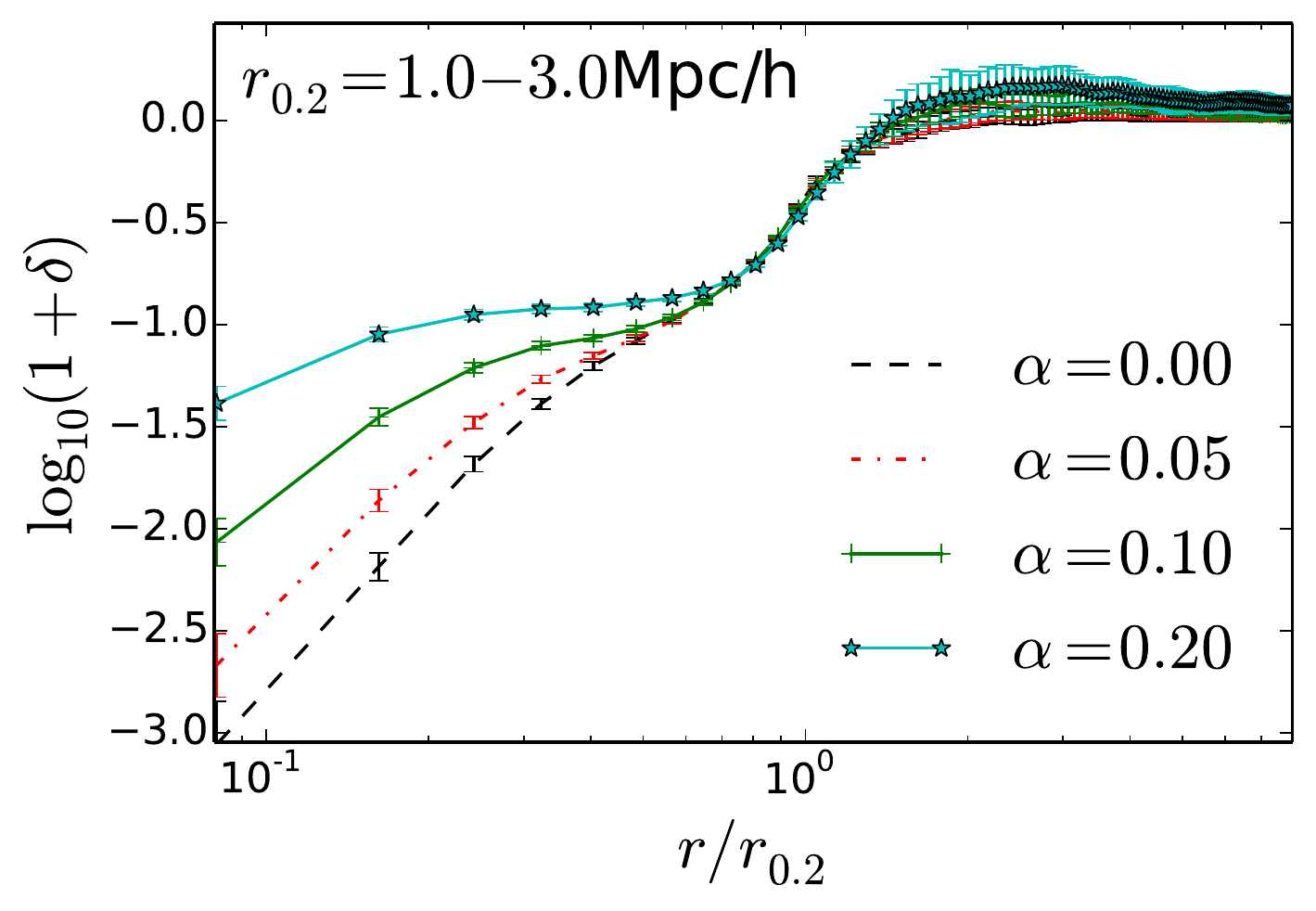}
    \caption{Void density profiles as in Fig.\ \ref{fig:densprof1}, except scaled with $r_\mathrm{0.2}$. The left panel shows the voids in the  $0.5-1.0$\hmpc\ radius bin; the right panel shows the $1.0-3.0$\hmpc\ bin.}
    \label{fig:densprof2}
 \end{figure*}

 \begin{figure*}
    \centering
     \includegraphics[width=0.45\textwidth]{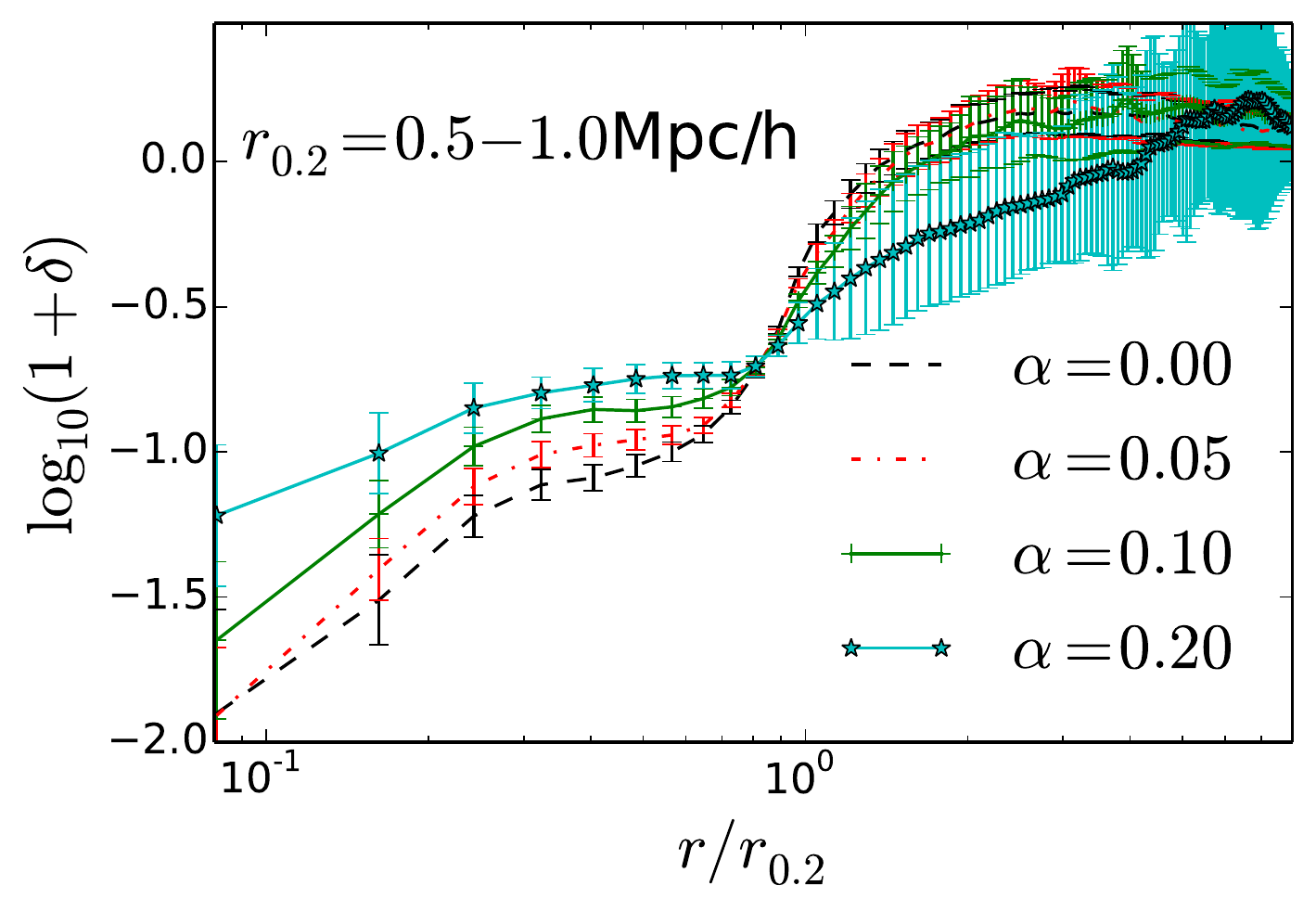}
     \includegraphics[width=0.45\textwidth]{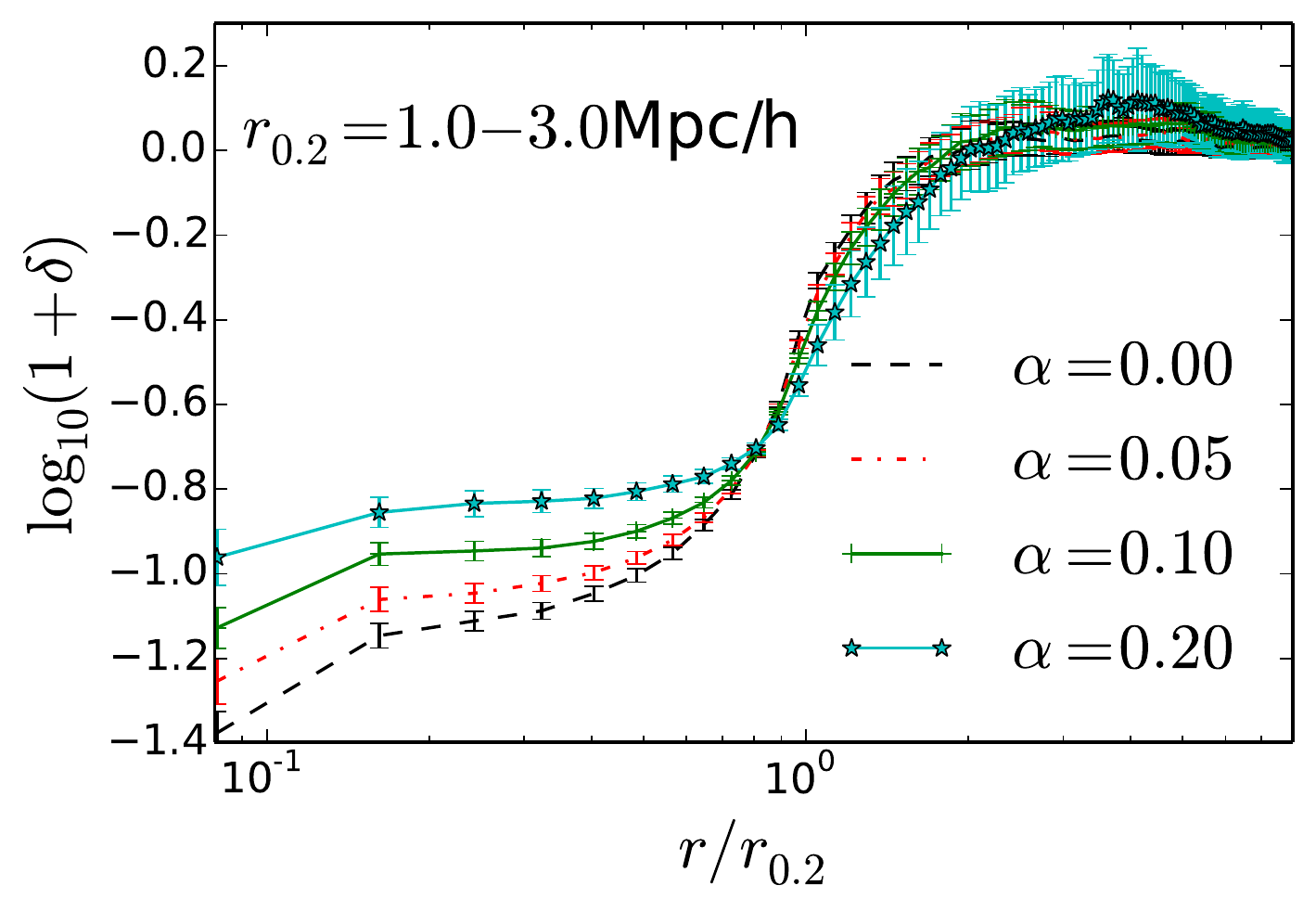}
    \caption{Void density profiles as in Fig.\ \ref{fig:densprof2}, except using void volume centroids as centers.}
    \label{fig:densprof3}
 \end{figure*}

We take special interest in void density profiles, as they have been measured by several different authors recently \citep{hamaus2014focus, clampitt2014lensing, 2014MNRAS.443.3238P,  nadathur2014self, 2014MNRAS.440..601R}. We show density profiles in Fig.\ \ref{fig:densprof1}.  As found by \citet{Hamaus:2014vy}, smaller voids are, on average, deeper. 

Our voids are mostly in the radius range 1-10 \hmpc. We divide the voids into two bins of effective radius, 1-5 and 5-10\hmpc.  We measure the density profiles { starting from void centers using linear radial bins. For radial bin $[r, r+\Delta r)$, the density is simply 
${3N_r}/{4\pi[(r+\Delta r)^3 - r^3]}$
where $N_r$ is the number of particles detected in this bin.   We investigate two definitions of the center: a) the actual density minimum of the void, as measured by the VTFE; and b) the volume centroid of the void, defined as $\sum{\mathbf{x}_iV_i}/\sum{V_i}$, where $\mathbf{x}_i$ is the position of particle $i$ belonging to the void, and $V_i$ is the Voronoi volume of that particle. The volume centroid would be easier to locate observationally than the density minimum. These profiles are further scaled by different radii  $r_s$ (i.e. $r_\mathrm{eff}$ or $r_{0.2}$) using linear interpolation.
Note that for the profiles starting from the density minimum, the Voronoi tessellation always guarantees a particle in the center, and hence a spike.  We remove the central  bin to remove such an artificial spike. Error bars of each data bin were measured using the standard deviation divided by $\sqrt{N}$, where $N$ is the number of profiles stacked. The error bars shown in this paper are all 2-$\sigma$ errors.}

Since the voids detected by \zobov\ are highly irregular in shape, \zobov's effective radius may not be the most meaningful radius measure in all cases. We visually check the voids and found that most of the irregularities are in the noisy edges of the voids.  We therefore define a radius $r_{0.2}$ for voids such that at this point the average density encompassed is $0.2\bar{\rho}$, following \citet{Jennings2013}. The distribution of $r_{0.2}$ is shown in Fig.\ \ref{fig:r02pdf}.   $r_{0.2}$ was typically from $0.5-3$\hmpc.  {The distribution of voids of $r_{0.2}\gtrsim 0.5$ \hmpc\ shows similar features as does $r_\mathrm{eff}$ due to the small scale suppression effect of WDM. Note that the number of small voids, with $r_{0.2}\lesssim 0.8$\hmpc, is curiously higher in WDM than in CDM. For these poorly-resolved small voids, suppose WDM voids are simply shallower versions of CDM voids. The WDM voids will tend to have smaller $r_{0.2}$, since, starting from a higher density minimum, a sphere needs not go out as far to reach an enclosed density of $0.2\bar{\rho}$.} Again we use the core particle as the center, scaled the void profiles measured by shell bins by $r_{0.2}$, and stacked them, shown in Fig.\ \ref{fig:densprof2}. The central densities are quite similar across different voids.
Note that $r_{0.2}$ is typically 2-3 times smaller than $r_\mathrm{eff}$. This is surprisingly large, but may be the effect of including the full, generally irregularly shaped, density ridges around voids. As discussed in \cite{Jennings2013}, choosing a different definition of void size simply moves voids among radius bins, but does not change their total abundance.

The shallowing of the density profile in central bins is entirely unsurprising if the profiles are measured from density minima; this follows almost trivially from the increase in density minima. Less obvious is the behaviour of void profiles as measured from their volume centroids. The profile from the volume centroid is more observationally relevant, because there is some hope of inferring it from a dense galaxy sample, while locating a 3D density minimum would be quite difficult.   In Fig.\ \ref{fig:densprof3}, we show density profiles measured in the same way as before, except from the volume centroids  using the Voronoi volumes of all particles reported in the \zobov\ void.
These profiles are scaled by $r_{0.2}$ and stacked in the same radius bins as in the previous case.  As previously, we tried to use $\reff$ to scale and stack, but  in highly noisy profiles. The reason is the same: the shapes of the voids become irregular when their sizes are small, and there is some randomness in whether part of a density ridge is included or not, so the \zobov\ effective radius can be noisy.

Either scaling the profile with $r_{0.2}$ or $\reff$, the density profiles show some universalities -- the central part of the profile is relatively stable. For different DM settings, the center part of the profiles is clearly different. While the profiles when scaling by $r_{0.2}$ are less { noisy} than those scaled by $\reff$, the profiles' shapes are similar in both cases.
It is reassuring that the results hold whether the volume centroid or the density minimum is used to measure the density profile.

In most previous profile studies, people use the effective radius $\reff$ to scale the density profile and get universal profiles -- a relatively flat central plateau, a sharp edge and a compensated wall, tending to unity faraway. We argue that using $r_{0.2}$, the radius at which the mean enclosed density reaches 0.2, as a scaling constant to find small void profiles is better. The void wall radius deviates from $\reff$ significantly with non-spherical shapes, while $r_{0.2}$ characterizes every void in the context of a spherical-evolution model. If the voids are self-similar, as stated in \cite{nadathur2014self}, $r_{0.2}$ surely returns more consistent profiles. This is indeed shown in Fig.\ \ref{fig:densprof2} and Fig.\ \ref{fig:densprof3}. Since the slope of profiles is apparently the largest at $r\approx r_{0.2}$, a lensing measurement would be most sensitive if using this definition.

\section{Conclusion and Discussion}
We measure statistics and density profiles of voids in different dark matter settings, namely CDM, and WDM with characteristic scales $\alpha=0.05$, $0.1$ and $0.2$ \hmpc. In summary, the voids in WDM are shallower.
{\textwarning } WDM voids also tend to be larger, although this effect depends somewhat on the void definition. The number of statistically significant voids is also smaller in WDM simulations. The main question this paper poses is whether void density profiles can be used to detect WDM, or some other process that attenuates initial small-scale power. Our answer is yes, in principle. 

{\textwarning One advantage of using voids rather than high-density regions is that the structure of voids is much less sensitive to baryonic physics than in high-density regions. This is because matter in voids undergoes no stream-crossing on cosmological scales \citep[e.g.][]{FalckNeyrinck2015}. Substantial differences between dark-matter and baryonic physics are only expected when streams collide; in collisionless dark matter, the streams pass through each other, while streams of gas collide, e.g.\ forming shocks.}

However, it is quite difficult to constrain matter density profiles observationally, even if the voids themselves can be located using galaxies or other tracers, which becomes tricky at small radius because of issues such as redshift-space distortions. Lensing could be used to constrain density profiles, as shown in \citep{clampitt2014lensing}. 
However, lensing is only sensitive to the gradient of the surface density, which becomes zero at the very centre of the void.
Fortunately, our results do show a difference in the density gradient in different WDM scenarios.  Another possible probe is the integrated Sachs-Wolfe effect \citep[e.g. ][]{GranettEtal2008}, usefully sensitive to the potential, although that measure is difficult to detect for small voids because of the dominant primordial CMB.  But there are other probes of the potential through voids, such as in fluctuations in the cosmic expansion rate as measured with supernovae.  Another aspect of voids that we did not measure, but would also be sensitive to matter density profiles, is the velocity field within voids; perhaps very faint tracers such as absorption lines can be used to constrain these \citep[e.g.][]{TejosEtal2012}.  Another possible direction to constrain WDM is through the properties of filaments and perhaps walls in WDM, which we quantitatively showed become more prominent; however, we leave a detailed study of their properties such as density profiles to future work. 

\section*{Acknowledgments}
We thank Alex Szalay for useful comments and for providing access to high-performance computers for the numerical simulations, and thank Cai Yanchuan and Xin Wang for helpful conversations. We thank the anonymous referee for suggestions that helped to improve this paper.
LFY and MN are grateful for support from a grant in Data-Intensive Science from the Gordon and Betty Moore and Alfred P. Sloan Foundations and from NSF grant OIA-1124403 and OCI-1040114. MN is thankful for support from a New Frontiers in Astronomy and Cosmology grant from the Sir John Templeton Foundation. The research of JS has also been supported at IAP by ERC project 267117 (DARK) hosted by Universit\'e Pierre et Marie Curie - Paris 6. BF acknowledges support from STFC grant ST/K00090/1.
\label{pageend}

\bibliography{ms}{}
\bibliographystyle{mn2e}
\end{document}